\documentstyle[preprint,revtex]{aps}
\begin{document}
\draft
\preprint{IPT-EPFL, 1993}
\begin{title}
Exact Results of the 1D $1/r^2$ Supersymmetric t-J Model
\\
without Translational Invariance
\end{title}
\author{C. Gruber and D. F. Wang}
\begin{instit}
Institut de Physique Th\'eorique\\
\'Ecole Polytechnique F\'ed\'erale de Lausanne\\
PHB-Ecublens, CH-1015 Lausanne-Switzerland.
\end{instit}
%%%%%%%%%%%%%%%%%%%%%%%%%%%%%%%%%%%%%%%%%%%%%%%%%%%%%%%%%%%%%%%%
\begin{abstract}
In this work, we continue the study of the supersymmetric
t-J model with $1/r^2$ hopping and exchange without translational
invariance. A set of Jastrow eigenfunctions are obtained
for the system, with eigenenergies explicitly calculated.
The ground state of the t-J model is included in this set of
wavefunctions. The spectrum of this t-J model consists
of equal-distant energy levels which are highly degenerate.
\end{abstract}

%%%%%%%%%%%%%%%%%%%%%%%%%%%%%%%%%%%%%%%%%%%%%%%%%%%%%%%%%%%%%%%%
\narrowtext
In recent years, there have been considerable interests in
study of low dimensional electronic models of strong
correlation, due to the possibility that the normal state
of the two dimensional novel superconductivity\cite{anderson} may share
some interesting feature of 1D interacting
electron system (non-Fermi liquid behavior).
The one-band two dimensional Hubbard model reduces to the
t-J model in the large on-site energy limit.
The Hubbard model and the t-J model have been under
intense study through various approaches.
For these strongly correlated electron models,
few exact results may be obtained in two dimensions.
The high dimensional versions are much harder to study
than their one dimensional ones.
In one dimension, however, the Bethe-ansatz technique may allow
us to exactly solve Hamiltonians in some special cases, such
as the Lieb-Wu solution\cite{lieb} and the
ordinary t-J model at its supersymmetric point\cite{sakar}.
In particular, the 1D long range exactly solvable electronic
models have attracted a lot of attention, since
they display interesting physics with solutions of simple
mathematical
%% FOLLOWING LINE CANNOT BE BROKEN BEFORE 80 CHAR
structure\cite{shas88,hald91,shastry,shastry2,vacek,kura91,kawakami92,kawakami93,gruber,wang,poly,fowler,frahm,suth72,rucken92,ha,coleman,wang3}.

Recently, we have been able to explicitly construct all the constants
of motion for the translational invariant
long range supersymmetric t-J model, by mapping
the system to a mixture of fermions and bosons, with
the super-algebra representation for the electron fields\cite{gruber}.
Moreover, we have introduced
a one dimensional supersymmetric t-J model with
$1/r^2$ hopping and exchange without translational invariance.
This system has also been shown by us to be completely integrable,
with infinite number of conserved quantities
explicitly constructed\cite{gruber}.
In this work, we continue the study of this t-J model. A set of
Jastrow eigenfunctions, as well as their eigenenergies, are obtained
explicitly. The ground state of the t-J model is included in this set
of wavefunctions. We also briefly discuss the structure of the full
spectrum for the system.

The Hamiltonian for the one-dimensional t-J model is given by\cite{gruber}
\begin{equation}
H_{tJ} =(1/2) P_{G}
\left[ -\sum_{1\le i\ne j \le L}\sum_{\sigma=1}^{N} t_{ij} ( c_{i\sigma}
^{\dagger}
c_{j\sigma}) +
\sum_{1\le i\ne j \le L}
J_{ij} \left[ P_{ij}-(1-n_i)(1-n_j)\right] \right] P_{G},
\label{eq:original}
\end{equation}
where we take the hopping matrix and the spin exchange interaction to be
$t_{ij}/2 = J_{ij} = 1/(r_i-r_j)^2$.
$L$ is the number of sites on the chain.
In this model,
the positions of the sites $\{ r_i \}$ are given by the roots of the
Hermite polynomial $H_L(x)$, and the spin component $\sigma$ takes
values from 1 to $N$.
The operator $c_{i\sigma}^{\dagger}$ is the
the creation operator for an electron at site $i$ with spin $\sigma$,
$c_{i\sigma}$ is the corresponding
annihilation operator. Their anti-commutation relations
are given by $\{c_{i\sigma_i},c_{j\sigma_j}^{\dagger}\}_+
=\delta_{ij} \delta_{\sigma_i\sigma_j},
\{c_{i\sigma_i},c_{j\sigma_j}\}_+=0, \{c_{i\sigma_i}^\dagger
,c_{j\sigma_j}^\dagger\}_+=0$.
The Gutzwiller projection operator $P_{G}$
projects out all the double or multiple occupancies,
$P_{G} = \prod_{i=1}^L P_{G}(i)$, and $P_{G}(i) =\delta_{0,n_i}
+\delta_{1,n_i}$, with $n_i=\sum_{\sigma=1}^N c_{i\sigma}^\dagger
c_{i\sigma}$.
The operator $P_{ij}=\sum_{\sigma=1}^N \sum_{\sigma'=1}^N
c_{i\sigma}^{\dagger} c_{i\sigma'} c_{j\sigma'}^{\dagger} c_{j\sigma}$
exchanges the spins of the electrons at sites $i$ and $j$, if both sites
are occupied, and is zero otherwise.
At half-filling, our t-J model becomes the long range
spin model, introduced first by Polychronakos
on such a non-translational-invariant
lattice\cite{poly}.

In terms of the $b$ and $f$ fields, the eigenequation of the t-J model
can be written as\cite{gruber}
\begin{equation}
- \sum_{1\le i < j \le  L} (q_i - q_j)^{-2} M_{ij}
\phi (\{q\}, \{\sigma\}) = E \phi (\{q\},\{\sigma\}),
\label{eq:eigenequation}
\end{equation}
where $\phi (\{q\},\{\sigma\}) = \phi (q_1 \sigma_1, q_2\sigma_2,
\cdots, q_{N_e}\sigma_{N_e} | q_{N_e+1}, q_{N_e+2}, \cdots, q_L)$
is the amplitude for the $f$ fermions to be at $q_1, q_2, \cdots, q_{N_e}$,
while the spinless $b$ bosons are at $q_{N_e+1}, q_{N_e+2}, \cdots, q_L$.
Here, $\{\sigma\}=(\sigma_1, \sigma_2, \cdots, \sigma_{N_e} )$ and
$\{q\}=(q_1, q_2, \cdots, q_L) = (x_1, x_2, \cdots, x_{N_e}, y_1, y_2,
\cdots, y_Q)$. The wavefunction $\phi$ is symmetric in the $b$ boson
positions $\{y\}$, while antisymmetric when exchanging $x_i\sigma_i$ and
$x_j\sigma_j$. The operator $M_{ij}$ exchanges only the position
variables $q_i$ and $q_j$, that is, $M_{ij} \phi (\{q\}, \{\sigma\})
=\phi (\{q'\}, \{\sigma\})$, with $\{q\}= (q_1, q_2,\cdots, q_i, \cdots, q_j,
\cdots, q_L)$ and $\{q'\}=(q_1, q_2, \cdots, q_j,\cdots,q_i,\cdots, q_L)$.
In this approach, the $f$ fermions and the $b$ bosons occupy the whole chain,
i.e. $\{q\}$ and $\{q'\}$ are permutations of the sites
$\{r_1,r_2,\cdots, r_L\}$, and
we work in the Hilbert space where at each site
there is exactly one particle. $Q$, the number of the $b$ bosons,
is also the number of holes in the original problem;
$N_e$, the number of the $f$ fermions, is also the number of
the $c$ electrons on the lattice. $\tilde N_{\alpha}$, with $\alpha
=1, 2, \cdots N$,
the number of the $f$ fermions with spin component $\alpha$,
is also the number of the $c$ electrons with spin component $\alpha$
on the chain.

The Hamiltonian in the first quantization, as given by
Eq.~(\ref{eq:eigenequation}), is
\begin{equation}
H= - \sum_{1\le i <j \le L} (q_i -q_j)^{-2} M_{ij}.
\end{equation}
It commutes with the permutation operator
$T_{ij} = P_{ij}^{\sigma} M_{ij}$ exchanging the $f$ fermion spin and
position simultaneously.
Let us work in the Hilbert space where the number of
fermions of each flavor is fixed,
i.e. $\tilde N_{\sigma}$, $\sigma =1, 2, \dots, N$, is fixed.
Consider the following wavefunction in Jastrow product form,
\begin{equation}
\phi (x_1\sigma_1, x_2\sigma_2, \cdots, x_{N_e}\sigma_{N_e}
|y_1, y_2, \cdots, y_Q)
=\prod_{i<j} (x_i-x_j)^{\delta_{\sigma_i\sigma_j}} e^{i{\pi\over 2}
sgn (\sigma_i -\sigma_j)},
\label{eq:groundstate}
\end{equation}
where $\{x\}$ and $\{y\}$ span the whole lattice.
We would like to show that this wavefunction
is an eigenstate of the system.
The Hamiltonian in Eq.~(\ref{eq:eigenequation})
can be broken up into three parts,
the first part $H_1$ exchanges the $f$ fermions,
the second $H_2$ exchanges the $b$ bosons, and the
third $H_3$ exchanges the bosons and the fermions:
\begin{eqnarray}
&&H_1=(-1) \sum_{1\le i<j\le N_e} (q_i-q_j)^{-2} M_{ij}\nonumber\\
&&H_2= (-1) \sum_{1+N_e \le \alpha<\beta  \le L} (q_\alpha-q_\beta)^{-2}
M_{\alpha\beta}\nonumber\\
&&H_3= (-1)
\sum_{N_e+1\le \alpha \le L} \sum_{1\le j\le N_e}
(q_\alpha -q_j)^{-2} M_{\alpha j}.
\end{eqnarray}
We then calculate the effects of these three parts when acting
the Jastrow wavefunction given by Eq.~(\ref{eq:groundstate}).
The contribution from $H_2$ is immediate:
\begin{equation}
H_2 \phi = -\sum_{\alpha < \beta }
( y_\alpha-y_\beta )^{-2} \phi.
\end{equation}
The contributions from $H_1$ and $H_3$ are harder to deal with
since many particle terms are involved. Using a similar trick
introduced in Refs.~\cite{vacek,ha}, we have
\begin{eqnarray}
&&H_3 \phi = -\sum_{i} \sum_{\alpha} (x_i -y_\alpha)^{-2}
\prod_{j (\ne i) } ({ y_\alpha -x_j \over x_i -x_j})^{
\delta_{\sigma_i\sigma_j}} \phi\nonumber\\
&& = - \sum_i \sum_{\alpha} (x_i -y_\alpha)^{-2} \prod_{j(\ne i)}
(1 + \delta_{\sigma_i\sigma_j} {y_\alpha -x_i \over x_i -x_j} ) \phi\nonumber\\
&&= - \sum_i \sum_{\alpha} \left[ (x_i -y_\alpha)^{-2} + \sum_{j(\ne i)}
{\delta_{\sigma_i \sigma_j} \over (y_\alpha -x_i ) (x_i -x_j)} \right]\phi
- Rest,
\end{eqnarray}
where
\begin{equation}
Rest = \sum_\alpha \sum_{\sigma=1}^N \sum_{r=3}^{N_e}
\sum_{J \subset \wp^\sigma \atop |J|=r} [ \sum_{i\in J} (y_\alpha -x_i )^{r-3}
\prod_{x_j \in X_J/x_i} {1\over (x_i-x_j)}  \phi ],
\end{equation}
with $\wp^\sigma = \{ k\in \{1, 2, \cdots, N_e\}; \sigma_k =\sigma\},
X_J =\{x_j; j\in J\}$. Then using the fact that for
any set $X = (x_1, x_2, \cdots, x_n)$, we have the
following identity ( see Appendix A )
\begin{equation}
\sum_{i=1}^n x_i^t \prod_{x_j \in X/x_i} {1\over (x_i -x_j)} =0,
\label{eq:sumrule}
\end{equation}
for all $t=0, 1, 2, \cdots, n-2$, we conclude that $Rest =0$.

The contribution from $H_1$ is calculated in a similar manner:
\begin{eqnarray}
&&H_1 \phi = -\sum_{i<j} (x_i -x_j)^{-2} (1-2\delta_{\sigma_i \sigma_j})
\prod_{k(\ne i, j)} ({x_k -x_j \over x_k -x_i })^{\delta_{\sigma_i\sigma_k}}
({x_k -x_i \over x_k -x_j })^{\delta_{\sigma_j\sigma_k}} \cdot \phi
\nonumber\\
&&=\sum_{i<j} \delta_{\sigma_i\sigma_j} (x_i -x_j)^{-2} \phi
-\sum_{i<j} (x_i -x_j)^{-2} (1-\delta_{\sigma_i\sigma_j})
\prod_{k(\ne i,j)}
(1 + \delta_{\sigma_i\sigma_k} {x_i -x_j\over x_k -x_i}
- \delta_{\sigma_j\sigma_k} {x_i -x_j \over x_k -x_j } )\nonumber\\
&&=\sum_{i \ne j} \delta_{\sigma_i\sigma_j} (x_i -x_j)^{-2} \phi
- \sum_{i<j} (x_i -x_j)^{-2}\phi  - \nonumber\\
&&- \sum_{i<j} (x_i -x_j)^{-2} (1-\delta_{\sigma_i\sigma_j})
\sum_{{\scriptstyle K_1\subset \{1, 2, \cdots, N_e\}/ij
\atop \scriptstyle  K_2\subset\{1,2,\cdots, N_e\}/ij}
\atop \scriptstyle K_1\cup K_2 \ne 0}
\prod_{k\in K_1} \delta_{\sigma_i\sigma_k} {x_i -x_j \over x_k -x_i }
\cdot \prod_{k\in K_2} {x_i -x_j \over x_k -x_j }
(-\delta_{\sigma_j\sigma_k}) \phi.
\end{eqnarray}
Using the sum rule Eq.~(\ref{eq:sumrule}), and the fact
that $\sum_{i\ne j \ne k} (x_k-x_i)^{-1} (x_k-x_j)^{-1}
\delta_{\sigma_i\sigma_j}\delta_{\sigma_i\sigma_k} =0$,
the last term in the above equation becomes
\begin{eqnarray}
&&-\phi \sum_{i<j } (1-\delta_{\sigma_i\sigma_j})
\sum_{k(\ne i,j)} \left[ (x_i-x_j)^{-1}
(x_k -x_i)^{-1} \delta_{\sigma_i\sigma_k}
- (x_i -x_j)^{-1} (x_k-x_j)^{-1}
\delta_{\sigma_j\sigma_k}\right] \nonumber\\
&&=-\phi \sum_{i\ne j} \sum_{k(\ne i,j)} (x_i-x_j)^{-1}
(x_k -x_i)^{-1} \delta_{\sigma_i\sigma_k}.
\end{eqnarray}
In the end, we have
\begin{equation}
H \phi = -\left[ \sum_{1\le i <j \le L} (q_i -q_j)^{-2} \right] \phi
+ \sum_{i\ne j} (x_i -x_j)^{-1} \delta_{\sigma_i\sigma_j}
\left[ \sum_{k(\ne i)}  (x_i -x_k)^{-1}
+ \sum_\alpha (x_i -y_\alpha)^{-1} \right] \phi.
\end{equation}
Using the properties of the roots of the Hermite polynomial
\begin{equation}
r_i = \sum_{j(\ne i)} (r_i -r_j)^{-1}, \sum_{1\le i <j \le L}
(r_i -r_j )^{-2} = L(L-1)/4,
\end{equation}
we thus conclude that the wavefunction $\phi$ is an eigenstate
with eigenvalue
\begin{equation}
E = -L(L-1)/4 + (1/2) \sum_{\sigma =1}^N (\tilde N_\sigma -1) \tilde N_\sigma.
\label{eq:eigenvalue1}
\end{equation}
Although it is expected that this wavefunction is the
lowest energy state in the subspace of fixed
$\tilde N_1, \tilde N_2, \cdots, \tilde N_N$, we were not able
to prove it. However, in the case of $SU(2)$, the small lattice
diagonalization up to 8 sites confirms this conjecture. Moreover,
the discussion below will also confirm this idea for the general case.
For fixed number $N_e$ of the electron number, the minimum of the
energy is obtained when $|\tilde N_\sigma - \tilde N_{\sigma'}|$
is as small as possible for each pair $\sigma \ne \sigma'$.

In the $SU(2)$ case, the above result becomes
\begin{equation}
E = (-1)L(L-1)/4 + (1/2) \tilde N_{\uparrow}  ( \tilde N_{\uparrow} -1)
+(1/2) \tilde N_{\downarrow} (\tilde N_{\downarrow}-1),
\label{eq:energy}
\end{equation}
where $\tilde N_{\uparrow}$ and $\tilde N_{\downarrow}$ are the numbers of
the up-spin electrons and the down-spin electrons respectively.
For fixed number of electrons on the chain, i.e. for fixed $N_e$,
the minimum of the energy given in Eq.~(\ref{eq:energy})
is obtained when $S_z=0$ for even $N_e$, or when
$S_z=\pm 1/2$ for odd $N_e$.
Therefore, the ground state
is a spin singlet ( respectively spin 1/2 ) state for even ( respectively odd)
number of electrons on the chain.
In particular, for an even number of electrons on the chain,
the ground state energy is
\begin{equation}
E_G= (-1/4) L(L-1) + N_e^2/4 -N_e/2,
\end{equation}
while for an odd number of electrons it is
\begin{equation}
E_G = (-1/4)L(L-1) +({N_e -1 \over 2})^2.
\end{equation}
The charge susceptibility of the ground state $\chi_c$ is given by
$\chi_c^{-1} =\partial^2  E_G/\partial N_e^2 = 1/2$, independent of the
electron concentration. Very unexpectedly,
the charge susceptibility is also finite at
half-filling $N_e=L$, in spite of the existence of
a metal-insulator phase transition at half-filling for this system.
This is in contrast to the case of the periodic $1/r^2$ supersymmetric
t-J model, where the charge
susceptibility is divergent at half-filling, at which the metal-insulator
phase transition occurs.

To study the spectrum of the system away from half-filling,
we follow the idea introduced in Ref.~\cite{poly}. Let us
define the operators
\begin{eqnarray}
&&\pi_j = i \sum_{k(\ne j)} (q_j-q_k)^{-1} M_{jk} = \pi_j^\dagger ,\nonumber\\
&&a_j^\dagger = \pi_j +i q_j,\nonumber\\
&&a_j = \pi_j -iq_j,
\end{eqnarray}
which satisfy the following commutation relations:
\begin{eqnarray}
&&[ \pi_j, \pi_k ] =0 \nonumber\\
&&[q_j, H] = i\pi_j\nonumber\\
&&[\pi_j, H] = -2 i \sum_{k(\ne j)} (q_j-q_k)^{-3}.
\end{eqnarray}
Then using the property of the roots of the Hermite polynomial
we have
\begin{equation}
[\pi_j, H]=-iq_j,
\end{equation}
and
\begin{eqnarray}
&& [ a_j^\dagger, H] = -a_j^\dagger\nonumber\\
&& [ a_j, H] =a_j.
\end{eqnarray}
Therefore the operators
$A_i^\dagger (\nu) = a_i^\dagger S_i^{(\nu)}, i=1, 2, \cdots, N_e$,
where $\nu =0,\pm, z$ for the $SU(2)$ case with $S_i^{(0)}=1$,
will act as raising operators,
while their hermitian conjugate $A_i(\nu)$
will act as lowering operators. It thus follows that the wavefunction
\begin{equation}
\phi_{\{n\}, \{\nu\}} = \sum_{P}
\prod_{i=1}^{N_e} (A_i^\dagger(\nu_i))^{n_{P_i}}\phi
\end{equation}
with $\{n\}=(n_1, n_2, \cdots, n_{N_e}),
n_i\ge 0, \{\nu\}=(\nu_1, \nu_2, \cdots,
\nu_{N_e} )$, is either an eigenstate with energy
\begin{equation}
E_{\{n\}} = E +\sum_{i=1}^{N_e} n_i
\end{equation}
or zero.

Moreover, it is shown in the Appendix B that the operators
$\sum_{i=1}^{N_e} A_i (\nu_i)$ with $\nu_i=0$ or $z$,
and $a_\alpha$ annihilate the wavefunction
$\phi$, and also $\sum_{i=1}^{N_e} A_i (\pm) \phi_G =0$,
confirming the conjecture that $\phi$ is the lowest energy state
in the subspace. We then arrive at the conjecture that the excitation
spectrum of the system is of the form
\begin{eqnarray}
E(s) = &&E + s; \nonumber\\
&&s\in (0, 1, 2, \cdots, s_{max}),
\label{eq:fullspectrum}
\end{eqnarray}
i. e., the spectrum of this t-J model
consists of equal-spaced energy levels. Since the model is on
a finite chain, $s_{max}$ is finite. In the special case
of $SU(2)$, the small lattice diagonalization up to 8 sites
suggests that the highest energy level is given by
\begin{equation}
E_{max} (Q) = L(L-1)/4 -Q(Q-1)/2,
\end{equation}
i.e., for an even number of electrons,
\begin{equation}
E_{max}=E_G +(1/4) N_e (4 L-3 N_e),
\end{equation}
where $Q$ is the number of holes on the chain, $N_e =L-Q$ is the
number of electrons. For $N_e =1$ or $N_e =L$ ( half -filling ),
this formular gives right results; moreover, at half-filling,
this corresponds to all spins polarized in one direction.

The feature of the t-J model spectrum consisting of equal-distant
energy levels may also be seen by taking the strong
interaction limit of the Sutherland-Calogero-Morse quantum system
for a mixture of fermions and bosons,
\begin{equation}
H = (-1/2) \sum_{i=1}^L \partial^2/\partial q_i^2 +
\sum_{i=1}^L l^2 q_i^2 /2 +
\sum_{i<j} l(l-M_{ij})/(q_i-q_j)^2,
\end{equation}
where there are $N_e$ fermions with spins and $Q$ spinless bosons,
$M_{ij}$ permutes the positions of the particles $i$ and $j$ only.
The mixture gas has equal-distant energy levels described in terms
of effective harmonic oscillators.
In the strong interaction limit, the elastic modes
decouple from the internal degrees of freedom. Since elastic modes
also consists of equal-distant energy levels, we thus are led
to the conclusion that the spectrum of the internal dynamics,
which is that of our t-J model,
also consists of equal-distant energy levels.
Further work is necessary for a fully complete proof
that the t-J model full spectrum takes the form Eq.~(\ref{eq:fullspectrum}).

Finally, we would like to point out that
the states of the t-J model in the whole Hilbert space are
grouped into a structure of ``spin supermultiplets'',
as indicated by the small lattice diagonalization, similar
to that of the periodic $1/r^2$ supersymmetric t-J model.
Such pattern of the Hilbert space is related to the
symmetries associated with the Hamiltonian. It is
highly worth while to identify them more explicitly, and
we would like to study these aspects in further work.

In summary, a set of Jastrow eigenfunctions have been found
for the t-J model, with the eigenenergies explicitly calculated.
The expected ground state of the t-J model is included in this set of
wavefunctions. The full spectrum of the t-J model is found to
have equal-distant energy levels which are highly degenerate.
It would be very interesting to understand the underlying
symmetry principles that give rise to such simple Hilbert
space structure. It remains to study various correlation functions,
as well as the thermodynamics, for this strongly correlated
electron system. It would also be very interesting to study
the effective field theory for the low-lying excitations for
this t-J model.

We wish to thank Dr. James T. Liu and Dr. Nicolas Macris for
conversations. In particular, we are very grateful to Dr. James T. Liu
for his substantial numerical support.
We also would like to thank the World Laboratory Foundation
for the financial support.

\newpage
%%%%%%%%%%%%%%%%%%%%%%%%%%%%%%%%%%%%%%%%%%%%%%%%%%%%%%%%%%%%%%%%%%%%%%%
\begin{appendix}{ \,}
In this appendix, we provide a brief proof for
the sum rule Eq.~(\ref{eq:sumrule})
for reader's convenience. The same argument can also be found
in previous works Refs.~\cite{vacek,ha}.
Let $X=(x_1, x_2, \cdots, x_n); X_i=X/x_i$, we wish to show
\begin{equation}
\sum_{i=1}^n x_i^t \prod_{x_j \in X_i} {1\over x_i -x_j} =0,
\forall t=0, 1, 2, \cdots, (n-2).
\end{equation}
Indeed, the Vandermonde determinant $V (X)$ has the property:
\begin{equation}
\prod_{j=1}^n (x-x_j)={V(Xx)\over V(X)},
\end{equation}
where $X=(x_1,x_2,\cdots, x_n), Xx=(x_1,x_2,\cdots,x_n,x)$.
Therefore we obtain
\begin{eqnarray}
&&\sum_{i=1}^n x_i^t \prod_{x_j \in X_i} {1\over x_i -x_j} =
\sum_{i=1}^n x_i^t {V(X_i)\over V(X_ix_i)} =\nonumber\\
&&=\sum_{i=1}^n (-1)^{i-n} x_i^t {V(X_i)\over V(X)}=
{(-1)^{n-1} \over V(X)} det \pmatrix{x_1^t&x_2^t&\cdots&x_n^t\cr
                           1&1&\cdots&1\cr
                           x_1&x_2&\cdots&x_n\cr
                           \vdots&\vdots&\cdots&\vdots\cr
                           x_1^{n-2}&x_2^{n-2}&\cdots&x_n^{n-2}\cr}=0.
\end{eqnarray}
This thus proves the sum rule Eq.~(\ref{eq:sumrule}).
\end{appendix}
%%%%%%%%%%%%%%%%%%%%%%%%%%%%%%%%%%%%%%%%%%%%%%%%%%%%%%%%%%%%%%%%%%%%%%%%%%%%
\begin{appendix}{\,}
In this appendix, we shall show that
the lowering operators $A_i(z)$ and $a_{\alpha}$ give
zero when acting on the Jastrow wavefunction $\phi$ given by
Eq.~(\ref{eq:groundstate}). This will yield
a partial confirmation that the wavefunction $\phi$ is the lowest energy
state in the subspace where the number of particles of each
spin component is fixed.
For the $b$ boson degrees of freedom, we
have following property
\begin{equation}
a_i \phi =0, i\in (N_e +1, N_e +2, \cdots, L),
\label{eq:holes}
\end{equation}
which is shown to be true using the sum rule Eq.~(\ref{eq:sumrule})
and the property of the
Hermite polynomial roots $\sum_{ j(\ne i) }(r_i-r_j)^{-1} = r_i$.
The procedure
to deal with the permutation operator $M_{ij}$ in $a_i$
is very similar to that of proving $\phi$ to be the eigenenergy state
of the Hamiltonian, but we do not write the full details here.
Combining the Eq.~(\ref{eq:holes}) with the fact $\sum_{i=1}^L a_i =0$,
we thus arrive at the following results:
\begin{eqnarray}
&&\sum_{\alpha =N_e +1}^L a_{\alpha} \phi =0,\nonumber\\
&&\sum_{i=1}^{N_e} a_i \phi =0.
\end{eqnarray}
Furthermore, we realize that
\begin{equation}
[\sum_{i=1}^{N_e} A_i(z) ] \phi =0.
\label{eq:anni}
\end{equation}
We have been able to show this to be true, following the similar
approach to handle the effect of the permutation operator $M_{ij}$
acting on the Jastrow wavefunction $\phi$.
In the particular case where $\tilde N_{\uparrow} = \tilde N_{\downarrow}$,
the wavefunction function $\phi$ is a spin singlet and
we may globally rotate the first identity of Eq.~(\ref{eq:anni}) in the
spin space, giving us
\begin{equation}
\left[ \sum_{i=1}^{N_e} (A_i(\pm)) \right] \phi = 0.
\end{equation}
In summary, we have proved that it is impossible to
construct non-vanishing eigenstates with the lowering operators
and the wavefunctions $\phi$ in the subspace
where the number of electrons of each flavor is fixed.
\end{appendix}
%%%%%%%%%%%%%%%%%%%%%%%%%%%%%%%%%%%%%%%%%%%%%%%%%%%%%%%%%%%%%%%%%%%%%%%%%%%%

\end{document}